\documentclass[prapplied,twocolumn, superscriptaddress]{revtex4-2}
\usepackage{xcolor,amsmath,url,graphicx,hyperref}
\usepackage[capitalise]{cleveref}

\makeatletter
\def\maketitle{
	\@author@finish
	\title@column\titleblock@produce
	\suppressfloats[t]}
\makeatother

\begin{document}

\title{Volume dependence of microwave induced excess quasiparticles in superconducting resonators} 

	\author{Steven A. H. de Rooij}
		\email{s.a.h.de.rooij@sron.nl}
		\affiliation{SRON - Netherlands Institute for Space Research, Niels Bohrweg 4, 2333 CA Leiden, The
			Netherlands}
		\affiliation{Faculty of Electrical Engineering, Mathematics and Computer Science, Delft University of
			Technology, Mekelweg 4, 2628 CD Delft, the Netherlands}

	\author{Jochem J. A. Baselmans}
		\affiliation{SRON - Netherlands Institute for Space Research, Niels Bohrweg 4, 2333 CA Leiden, The
			Netherlands}
		\affiliation{Faculty of Electrical Engineering, Mathematics and Computer Science, Delft University of
			Technology, Mekelweg 4, 2628 CD Delft, the Netherlands}
			
	\author{Juan Bueno}
		\affiliation{Faculty of Electrical Engineering, Mathematics and Computer Science, Delft University of
			Technology, Mekelweg 4, 2628 CD Delft, the Netherlands}
	
	\author{Vignesh Murugesan}
		\affiliation{SRON - Netherlands Institute for Space Research, Niels Bohrweg 4, 2333 CA Leiden, The
			Netherlands}
	
	\author{David J. Thoen}
		\affiliation{SRON - Netherlands Institute for Space Research, Niels Bohrweg 4, 2333 CA Leiden, The
			Netherlands}
	
	\author{Pieter J. de Visser}
		\email{p.j.de.visser@sron.nl}
		\affiliation{SRON - Netherlands Institute for Space Research, Niels Bohrweg 4, 2333 CA Leiden, The
			Netherlands}

\date{\today}
\begin{abstract}
	The presence of quasiparticles typically degrades the performance of superconducting microwave circuits. The readout signal can generate non-equilibrium quasiparticles, which lead to excess microwave loss and decoherence. To understand this effect quantitatively, we measure quasiparticle fluctuations and extract the quasiparticle density across different temperatures, readout powers, and resonator volumes. We find that microwave power generates a higher quasiparticle density as the active resonator volume is reduced and show that this effect sets a sensitivity limit on kinetic inductance detectors. We compare our results with theoretical models of direct microwave photon absorption by quasiparticles and conclude that an unknown, indirect mechanism plays a dominant role in quasiparticle generation. These results provide a route to mitigate quasiparticle generation due to readout power in superconducting devices.
\end{abstract}

\maketitle 

\section{Introduction}
Superconducting devices have found numerous applications, ranging from advanced quantum circuits to the most sensitive radiation detectors. Quasiparticles (i.e., broken Cooper pairs) generally degrade the performance of these devices as they introduce microwave loss, decoherence \cite{Glazman2021} and reduced detector sensitivity \cite{deVisser2012}. The quasiparticle density exponentially decreases with decreasing bath temperature, but is typically observed to saturate at low temperatures \cite{Martinis2009,deVisser2011}. Multiple sources of quasiparticle generation have been identified and mitigated, including cosmic rays \cite{Martinis2021,McEwen2022}, radioactivity \cite{Cardani2021}, and stray light \cite{Baselmans2012a}.\\
We focus on the quasiparticle generation by the microwave signal used to read out the superconducting devices \cite{deVisser2012a}. The on-chip power of this readout signal ($P_{read}$) induces two nonlinear effects. The first is a nonlinear kinetic inductance effect. The kinetic inductance increases quadratically when the current becomes a significant fraction of the critical current. This nonlinear inductance is used in parametric amplifiers \cite{Esposito2021a} and tunable resonators \cite{Vissers2015}. It is an effect of the acceleration of the Cooper pair condensate. The second effect is the absorption of microwave photons by quasiparticles \cite{Eliashberg1972, Ivlev1973}, resulting in a redistribution to higher energies \cite{Goldie2012,deVisser2014b}. At low temperatures, this can lead to an increased quasiparticle density. Direct pair-breaking by microwave photons is not possible because their energy ($\hbar\omega_0$) is much smaller than the gap energy ($2\Delta$). However, when the quasiparticle relaxation rate is small compared to the photon absorption rate, quasiparticles can absorb multiple microwave photons. As a result, redistributed quasiparticles with energies $>3\Delta$ can emit phonons of $>2\Delta$ to break Cooper pairs. This leads to an excess quasiparticle density at low bath temperatures, which increases with $P_{read}$ \cite{deVisser2012a}.\\
Superconducting resonators are particularly sensitive to both these effects, since $P_{read}$ accumulates to an internal power, $P_{int}\propto Q P_{read}$, where $Q$ is the quality factor. The quality factor is typically limited by the coupling quality factor, $Q_c$, which is usually greater than $10^4$. Resonators are widely used in quantum circuits and as ultrasensitive radiation detectors spanning millimeter-wave to visible wavelengths, known as Microwave Kinetic Inductance Detectors (MKIDs) \cite{Day2003}. To reduce Two Level System noise (TLS), MKIDs are typically operated at a high $P_{int}$ \cite{Gao2007}, just below the bifurcation power, $P_{int}^{bif}$. At $P_{int}^{bif}$ the nonlinear kinetic inductance causes hysteresis in the resonance curve \cite{Swenson2013}. At such high $P_{int}$, we expect quasiparticle redistribution effects to become important as well.\\  
Another common approach to improve the sensitivity of MKIDs is reducing the absorption volume \cite{Hailey-Dunsheath2021}. However, in Ref. \cite{Baselmans2022} it is observed that decreasing $V$ by a factor 7 does not increase the sensitivity when driving close to $P_{int}^{bif}$. At the same time, the quasiparticle recombination time is observed to decrease with decreasing volume. That points towards a volume-dependent excess quasiparticle density, since the quasiparticle lifetime and density are inversely related. This raises the question: how are the excess quasiparticle density, MKID sensitivity, readout power and the resonator volume interrelated?

\begin{figure*}
	\includegraphics[width=\textwidth]{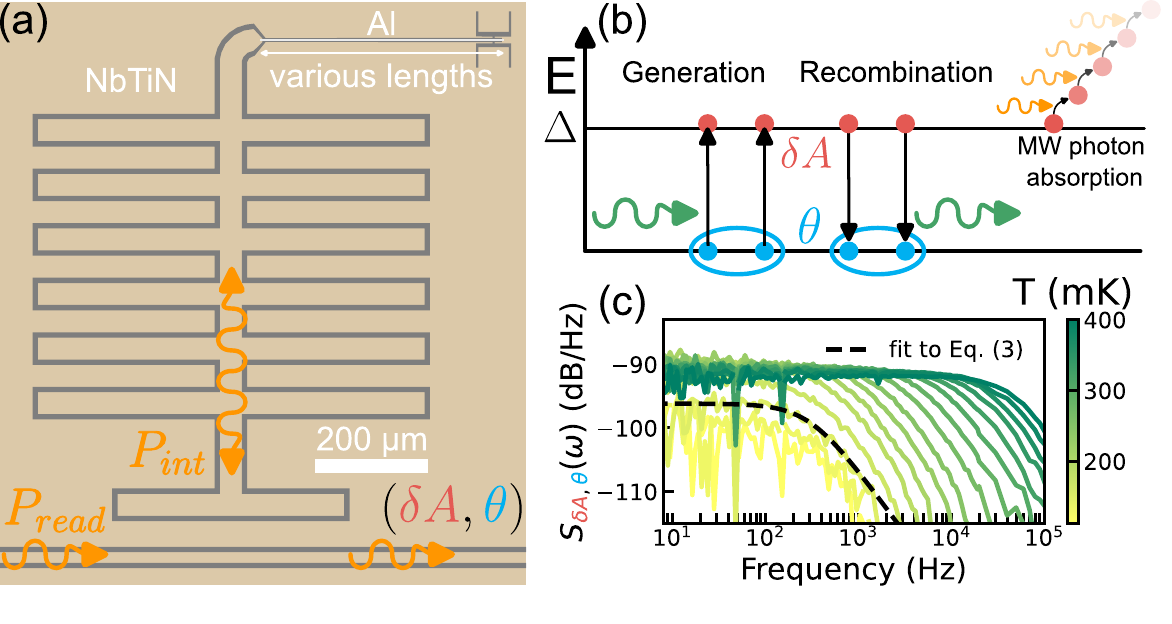}
	\caption{\label{fig:methods} Generation-recombination noise measurement on hybrid superconducting resonators. (a) Diagram of one of the resonators studied. The central line of the inductive part is a narrow ($1.7~\mu\text{m}$) Al line of various lengths. At the end of the inductive section, a twin-slot antenna is patterned in the NbTiN, which is not used in the measurement. The gaps to the NbTiN ground plane are 2 $\mu\text{m}$ for the inductive section. The NbTiN shunt capacitive part has a $40~\mu\text{m}$ wide central line and $8~\mu\text{m}$ wide gaps. We measure the forward transmission and calibrate it to the resonance curve to obtain $\delta A$ and $\theta$. (b) Energy diagram of the relevant processes regarding quasiparticles (red), Cooper pairs (blue), phonons (green) and microwave photons (orange). By analysing correlations in $\delta A$ and $\theta$, we extract information about the quasiparticle generation-recombination process. (c) Measured cross power spectral densities for one of the resonators, at different bath temperatures. From a fit (dashed line), we obtain the quasiparticle density and effective lifetime.}
\end{figure*}
\begin{figure*}
	\includegraphics[width=\textwidth]{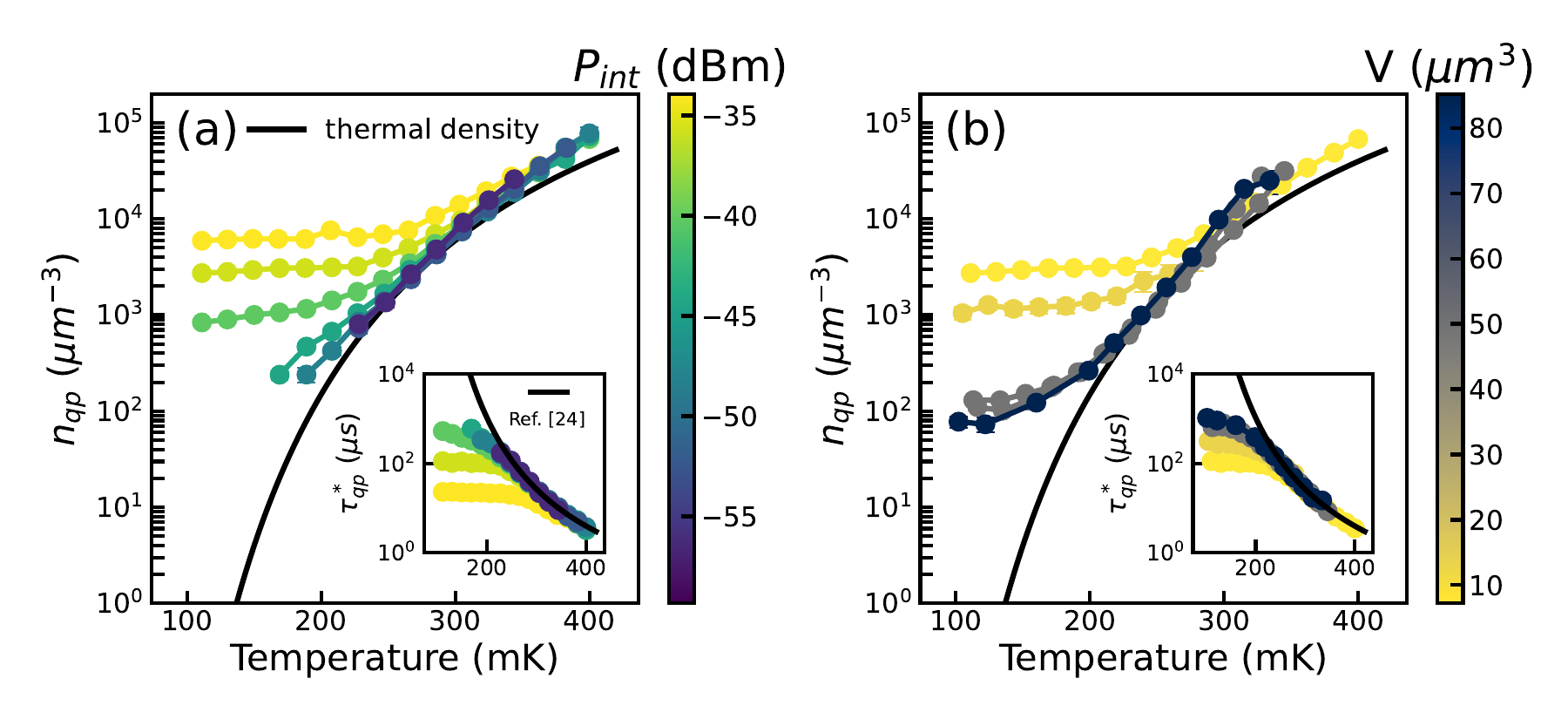}
	\caption{\label{fig:results} Results of the noise measurements for (a) the smallest Al volume resonator at various internal signal powers and (b) for various Al volumes when the microwave signal power is close to the bifurcation point. Error bars indicate statistical fit errors (see \cref{fig:nqpsat}(c)) and are smaller than the data points for most data points. The insets show the results for the effective quasiparticle lifetime, obtained from the same measurements. The black line in the insets is a fit to $\tau_{qp}^*(T)=\tau_{qp}(T)(1+\tau_{esc}/\tau_{pb})/2$, with $\tau_{esc}=0.42\pm0.07~\text{ns}$ as only free fit parameter. The error indicates the spread in fit results for the different resonators. We get $\tau_{pb}$ and $\tau_{qp}(T)$ for Al from \cite{Kaplan1976}. }
\end{figure*}
\section{Device design and measurement}
To answer this question, we measure the quasiparticle density in Al sections of different volumes, which are embedded in a NbTiN microwave resonator, see \cref{fig:methods}(a). The NbTiN (normal state resistivity $\rho_N=158~\mu\Omega\text{cm}$, critical temperature $T_c=14.6~\text{K}$, thickness $d=110~\text{nm}$) \cite{Thoen2017} is sputtered on a c-Sapphire substrate and patterned using a reactive ion etch as a large shunt interdigitated capacitor (IDC) \cite{Noroozian2009} to decrease the impedance of that section. By changing the length of the IDC fingers we set the resonance frequency, $\omega_0/(2\pi)$, between $4.5$ and $6.5~\text{GHz}$. The resonator is capacitively coupled to the transmission line with a coupling bar, which sets $Q_c$ between 30,000 and 100,000 for the resonators studied here. The inductive section is a co-planar waveguide with a sputtered and wet etched Al central line ($\rho_N=0.7~\mu\Omega\text{cm}$, $T_c=1.18~\text{K}$, $d=40~\text{nm}$), which is shorted at the end to make a quarter wave resonator.

In our analysis, we assume that we effectively probe an Al resonator with a volume equal to the volume of the Al strip. This is justified by the fact that the sensitivity to quasiparticles in the Al is significantly greater than for quasiparticles in the NbTiN, since the current density in the Al is uniform and much greater than in the NbTiN. The quasiparticles in the Al are confined in the strip, because of the larger gap energy of NbTiN. Furthermore, the microwave photons in the resonator mainly interact with the quasiparticles in the Al, because of the high impedance of the Al line and small Al volume compared to the NbTiN volume \cite{Fischer2023}. See \cref{ap:QPsens} for an estimate of the quasiparticle sensitivity and photon interaction in the Al versus the NbTiN.\\
We probe the resonator at its resonance frequency with $P_{read}$, resulting in an internal power \cite{Gao2008a}, 
\begin{equation}
	P_{int}=\frac{\hbar\omega_0^2}{\pi}\left<n_{ph}\right>=\frac{2 Q^2}{\pi Q_c}P_{read},
\end{equation}
where $\left<n_{ph}\right>$ represents the average number of photons in the resonator. $Q=(1/Q_c + 1/Q_i)^{-1}$ the loaded quality factor, with $Q_i$ the internal quality factor. We measure the forward transmission and calibrate it with respect to the resonance circle in the IQ-plane, to obtain the normalized amplitude ($\delta A$) and phase ($\theta$) \cite{Barends2008}. $\delta A$ corresponds to changes in the dissipation (i.e. $\delta (1/Q_i)$), induced by the quasiparticles, and $\theta$ to changes in  the kinetic inductance (i.e. $\delta\omega_0$), induced by Cooper pairs, see \cref{fig:methods}(b). We record noise in $\delta A$ and $\theta$ at different bath temperatures and calculate the cross power spectral density, shown in \cref{fig:methods}(c) \cite{deVisser2012a}. As we select for correlated changes in $\delta A$ and $\theta$, we effectively probe the generation-recombination process \cite{deVisser2012a}. This process is characterized by a Lorentzian spectrum \cite{deVisser2011,Wilson2004},  
\begin{equation}\label{eq:PSD}
	S_{\delta A, \theta}(\omega) = \frac{4\left<\delta N_{qp}^2\right>\tau_{qp}^*}{1 + (\omega\tau_{qp}^*)^2} \left[\frac{d \delta A}{dN_{qp}}\frac{d\theta}{dN_{qp}}\right].
\end{equation} 
Here, $N_{qp}$ is the average number of quasiparticles inside the Al volume and $\tau_{qp}^*$ is the effective quasiparticle recombination time. We measure the term inside the brackets, i.e. the amplitude and phase responsivities to changes in quasiparticle number, independently, by sweeping the bath temperature for $T>T_c/5$ and recording the change in resonance frequency ($\omega_0$) and internal quality factor ($Q_i$) \cite{deVisser2014}. From the same measurement, we obtain the kinetic inductance fraction, $\alpha_k$ \cite{Gao2006}, which is between 2 and 8\% for the devices studied here. With the responsivities known, we fit \cref{eq:PSD} to the data to obtain $\tau_{qp}^*$ and $\left<\delta N_{qp}^2\right>$. An example fit is shown in \cref{fig:methods}(c).\\
The quasiparticle fluctuation variance is given by $\left<\delta N_{qp}^2\right>=2\bar{R}N_{qp}^2\tau_{qp}^*/V$ \cite{Fischer2024a}, where $\bar{R}=2\Delta^2/((k_BT_c)^3N_0\bar{\tau}_0)$ is the recombination constant renormalized for phonon trapping (indicated by the bar). Here, $k_B$ is the Boltzmann constant and $N_0=1.72\times10^{4}~\mu\text{eV}^{-1}\mu\text{m}^{-3}$ is the single spin density of states at the Fermi level for Al \cite{Kaplan1976}. For $\Delta$, we use the BCS result, $2\Delta=3.52k_BT_c$. $\bar{\tau}_0=\tau_0(1+\tau_{esc}/\tau_{pb})$ is the renormalized electron-phonon time constant, with $\tau_0=438~\text{ns}$ describing the electron-phonon coupling strength and $\tau_{pb}=0.28~\text{ns}$ the phonon pair-breaking time in Al \cite{Kaplan1976}. $\tau_{esc}$ is the phonon escape time, which we get from a fit to the high temperature data of $\tau_{qp}^*=\tau_{qp}(1 + \tau_{esc}/\tau_{pb})/2$, with $\tau_{qp}=1/(2Rn_{qp}^T)$, where $n_{qp}^T$ is the thermal quasiparticle density (black line in \cref{fig:results}(a)). The fit is shown in the insets of \cref{fig:results} and results in $\tau_{esc}=0.42\pm0.07~\text{ns}$. From an analytical calculation of the phonon transparency of the Al-Sapphire interface \cite{Kaplan1979, Eisenmenger1976} we find $\tau_{esc}=0.40~\text{ns}$, which is consistent. Combining these results, we can extract the quasiparticle density, $n_{qp}=N_{qp}/V$ from the measured variance.\\
The expression we use for the variance has been recently calculated in Ref. \cite{Fischer2024a} and includes the effects of quasiparticle generation by microwave readout power. When phonons dominate the quasiparticle generation, it results in $\left<N_{qp}^2\right>=N_{qp}$, because in this regime $\tau_{qp}^*=1/(2\bar{R}n_{qp})$ \cite{Wilson2004}. When microwave power dominates, however, an excess quasiparticle density absorbs more microwave power, which results in a higher variance of $\left<N_{qp}^2\right>=2N_{qp}$, and $\tau_{qp}^*=1/(\bar{R}n_{qp})$.

\section{Results and discussion}
\cref{fig:results}(a) shows $n_{qp}$ obtained in this way, for one resonator and various $P_{int}$. At high temperatures $n_{qp}$ is equal to the thermal quasiparticle density, which verifies the method we use to extract $n_{qp}$. At high $P_{int}$ and low temperatures, $n_{qp}$ saturates at a value that increases with increasing $P_{int}$. The inset shows the results for $\tau_{qp}^*$ from the same fits, which saturates for low temperatures at lower values for higher $P_{int}$. This aligns with Ref. \cite{deVisser2012a}, confirming that the microwave power induces excess quasiparticles.\\
\cref{fig:results}(b) shows $n_{qp}$ for various Al volumes when $P_{int}$ is close to $P_{int}^{bif}$, i.e. in the high microwave power regime. We estimate the bifurcation power for the kinetic inductance nonlinearity as, 
\begin{equation}\label{eq:Pbif}
	P_{int}^{bif} = 0.64 N_0\Delta^2\frac{V\omega_0}{Q\alpha_k^2}.
\end{equation}
This expression is derived in \cref{ap:Pbif} for a quarter wave resonator with uniform current density.\\ 
We observe from \cref{fig:results}(b) that the excess quasiparticle density is higher and the effective quasiparticle lifetime decreases for the smaller Al volumes. We conclude that the quasiparticle generation by microwave readout power is more effective at small Al volumes.\\

\begin{figure*}
	\includegraphics[width=\textwidth]{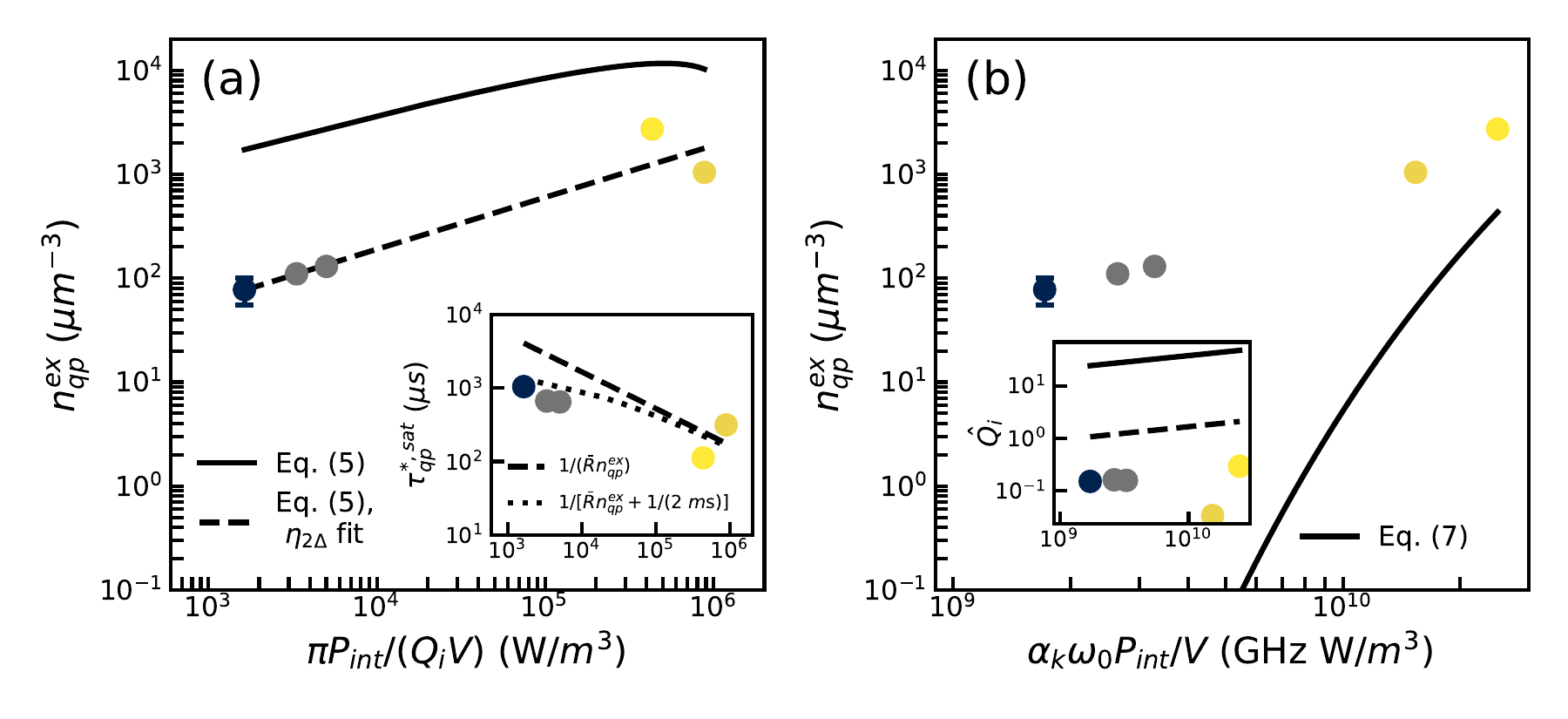}
	\caption{\label{fig:nqpsat} Excess quasiparticle density due to microwave power compared to kinetic equation calculations. (a) Comparison with the numerical study of Ref.\cite{Goldie2012} (\cref{eq:nqpex}). The solid line is \cref{eq:nqpex} with $\eta_{2\Delta}$ from Ref. \cite{Guruswamy2015} for Al. The dashed line is with $\eta_{2\Delta}=4\times10^{-4}$ as a fit parameter. The data points are the low temperature data of \cref{fig:results}(b) using the same colour coding for the device volume and with the error bars given by the standard deviation of the 3 lowest temperature points. The x-axis is $P_{abs}/V$ (see \cref{eq:Pabs}) with $Q_i$ from measurement. The inset shows the saturated effective quasiparticle lifetime from \cref{fig:results}(b) on the same x-axis as the main panel, compared to the expected effective lifetime from $\tau_{qp}^*=1/(\bar{R}n_{qp}^{ex})$ with $\eta_{2\Delta}=4\times10^{-4}$ as the dashed line. The dotted line includes an additional relaxation channel with a time constant of $2~\text{ms}$. (b) Comparison with the analytical description in Ref. \cite{Fischer2023} (\cref{eq:nqpsatFC}) without fit parameter. The main difference is that in Ref. \cite{Fischer2023}, the authors use an analytical expression for the quasiparticle-photon coupling constant, allowing $P_{int}$ as parameter, as opposed to $P_{abs}^{qp}$ in Ref. \cite{Goldie2012}. The data points are the same as in (a), but on a different x-axis. The inset shows the internal quality factor normalized to the quasiparticle density and resonator properties, $\hat{Q_i}=Q_i\alpha_k\hbar\omega_0n_{qp}^{ex}/(2N_0\Delta^2)$, on the same x-axis as the main panel. The dots are measured values, including $Q_i$ from a separate measurement and $n_{qp}^{ex}$ from the main panel. The solid line is the prediction for the quasiparticle creation regime and the dashed line for the redistribution regime, both from Ref. \cite{Fischer2023}. As the measured values are consistently lower, a process different from quasiparticle dissipation is likely to limit $Q_i$.}
\end{figure*}
\subsection{Excess quasiparticles due to dissipated microwave power}
To quantify this statement, we compare these results to theory. In Ref. \cite{Goldie2012}, the authors numerically study the kinetic equations describing the phonon and quasiparticle distribution functions in a non-equilibrium state \cite{Chang1977a}. The readout power effects are included via a microwave drive term \cite{Eliashberg1972, Ivlev1973}, which is proportional to the quasiparticle-photon coupling constant $c_{phot}^{qp}$. The authors of Ref. \cite{Goldie2012} implicitly determine $c_{phot}^{qp}$ by equating the absorbed power by the quasiparticles ($P_{abs}^{qp}$) to the power lost due to inelastic scattering (both quasiparticle-phonon scattering and recombination). They take $P_{abs}^{qp}$ as input parameter in the model. The dissipated power density in a resonator can be calculated as \cite{Gao2008a}
\begin{equation}\label{eq:Pabs}
	\frac{P_{abs}}{V} = \frac{2Q^2}{Q_i Q_c} \frac{P_{read}}{V}=\frac{\pi P_{int}}{Q_i V}.
\end{equation}
If $Q_i$ is dominated by quasiparticle losses (i.e. $Q_i=Q_i^{qp}$), $P_{abs}^{qp}$ is equal to $P_{abs}$ and can be obtained from measurements. The power lost due to inelastic scattering is obtained from the numerical calculations.\\
This results in two regimes. At high bath temperatures, thermal quasiparticles are redistributed to higher energies due to $P_{abs}^{qp}$, away from $\Delta$, and $Q_i$ increases ('redistribution regime'). At low bath temperatures, $P_{abs}^{qp}$ induces excess quasiparticles and $Q_i$ decreases ('quasiparticle creation regime'). These two regimes are experimentally observed in Ref. \cite{deVisser2014b}. The dependence of $\omega_0$ on $P_{int}$ was observed to be much stronger than expected from the excess quasiparticle density \cite{deVisser2014b}. This was later identified as a nonlinear kinetic inductance effect \cite{Semenov2016,Semenov2020a}.\\
Here, we focus on the quasiparticle creation regime. In \cref{fig:nqpsat}(a), we show the excess quasiparticle density, $n_{qp}^{ex} = n_{qp} - n_{qp}^T$, from \cref{fig:results}(b) as a function of the absorbed power density ($P_{abs}/V$, see \cref{eq:Pabs}).\\

In the quasiparticle creation regime, the excess quasiparticle density is predicted to be \cite{Goldie2012}, 
\begin{equation}\label{eq:nqpex}
	n_{qp}^{ex} = \sqrt{\eta_{2\Delta} \frac{\pi P_{int}}{Q_iV}\frac{8(1 + \tau_{esc}/\tau_{pb}) N_0^2\Delta k_B}{\Sigma_s}},
\end{equation}
where $\Sigma_s=3.4\times10^{10}~\text{W/m}^3/\text{K}$ for Al \cite{Goldie2012}, which describes the power flow density from the quasiparticles to the phonons. If only recombination (and no phonon scattering) would be taken into account, this would equal $\Sigma_s^{rec}=2\pi k_B (2N_0 \Delta)^2 R = 2.7\times10^{10}~\text{W/m}^3/\text{K}$, see \cref{ap:Pbif} for a derivation. The parameter $\eta_{2\Delta}$ in \cref{eq:nqpex} is the fraction of the phonons with an energy $>2\Delta$, which can be interpreted as a pair-breaking efficiency for the absorbed microwave power in the quasiparticle system. It has been numerically calculated in Ref. \cite{Guruswamy2015}. For Al in our $P_{abs}/V$ regime we get $\eta_{2\Delta}=0.20-0.013$, which results in the black line in \cref{fig:nqpsat}(a). This prediction clearly does not describe the measured quasiparticle density. If we use $\eta_{2\Delta}$ as fit parameter, we obtain $\eta_{2\Delta}=4\times10^{-4}$, shown by the dashed line in \cref{fig:nqpsat}(a). This value for $\eta_{2\Delta}$ is comparable to the values found in Ref. \cite{deVisser2012a} (where this parameter is defined as $\eta_{read}$). This description satisfactorily explains the data and we thus conclude that the excess quasiparticle density follows the dependence $n_{qp}^{ex}\propto \sqrt{P_{abs}/V}$.\\
This dependence results in a quasiparticle lifetime scaling as, $\tau_{qp}^*\propto1/n_{qp}\propto\sqrt{P_{abs}/V}$. We verify this in the inset of \cref{fig:nqpsat}(a). The dashed line is given by $1/(\bar{R}n_{qp}^{ex})$, with $n_{qp}^{ex}$ from \cref{eq:nqpex} with $\eta_{2\Delta}=4\times10^{-4}$. The deviation from the dashed line at low $P_{abs}/V$ could be due to additional quasiparticle relaxation channels, such as recombination in impurities \cite{Kozorezov2008,Barends2009a} or disorder-induced gap inhomogeneities \cite{deRooij2021,deRooij2024}. To account for that, we included a saturation lifetime of $2~\text{ms}$ in the dotted line, which is a typical value for Al \cite{Barends2009a}. The dotted line describes the data well, which supports our conclusion that $n_{qp}^{ex}\propto \sqrt{P_{abs}/V}$.\\

\subsection{Limiting MKID sensitivity}
The sensitivity of a power-integrating MKID is measured as the Noise Equivalent Power (NEP). This is ultimately limited by the fluctuations of quasiparticles in the absorber volume \cite{deVisser2012}, $\text{NEP}_{GR}=2\Delta/\eta_{pb}\sqrt{n_{qp} V/\tau_{qp}^*}$, where $\eta_{pb}$ is the pair-breaking efficiency of pair-breaking photons \cite{Guruswamy2014}. In the quasiparticle creation regime, this is thus limited by the excess quasiparticles generated by the microwave probe. To suppress other noise sources such as TLS and amplifier noise, typically a high probe power is chosen. If we assume the internal microwave power to be the maximum before bifurcation, $P_{int}=P_{int}^{bif}$ (\cref{eq:Pbif}), the internal quality factor is limited by quasiparticles, $Q_i=Q_i^{qp}$, and $k_BT\ll(\Delta, \hbar\omega_0)$, we can estimate this to be (see \cref{ap:NEP}), 
\begin{equation}
	\label{eq:NEP}
	\text{NEP}_{GR}^{bif} = 0.3\frac{\eta_{2\Delta}}{\eta_{pb}}\sqrt{\frac{\Delta^3}{\hbar\bar{R}}} \frac{\sqrt{V\omega_0}}{Q \alpha_k}.
\end{equation}
Filling in $\eta_{2\Delta}=4\times10^{-4}$, $\eta_{pb}=0.37$ \cite{Baselmans2022,Guruswamy2014} and the material and resonator parameters from Ref. \cite{Baselmans2022}, we obtain $1.8\times10^{-20}~\text{W}/\sqrt{\text{Hz}}$. Ref. \cite{Baselmans2022} measured a similar value $3.1\times10^{-20}~\text{W}/\sqrt{\text{Hz}}$ for all resonators, while the design parameters ($V$, $\omega_0$, $Q$ and $\alpha_k$) vary significantly. According to \cref{eq:NEP} the quasiparticle-limited NEP should vary as $\sqrt{V\omega_0}/(Q\alpha_k)$. This factor is approximately constant for all resonators in Ref. \cite{Baselmans2022}, which suggests that fluctuations of the microwave-induced excess quasiparticles limit the NEP of small volume MKIDs driven at high readout powers. This limit can be relaxed by decreasing the factor $\sqrt{V\omega_0}/(Q\alpha_k)$ in the MKID design. However, this will decrease the maximum internal power, since $P_{int}^{bif}\propto V \omega_0/(Q\alpha_k^2)$ (\cref{eq:Pbif}, and in turn increase TLS noise \cite{Gao2008b}. Therefore, to improve the NEP of MKIDs further, both the factor $\sqrt{V\omega_0}/(Q\alpha_k)$ and TLS noise must be reduced simultaneously.\\

\subsection{Microscopic mechanism for excess quasiparticles}
From \cref{fig:nqpsat}(a), we concluded that $P_{abs}/V$ generates excess quasiparticles with an efficiency $\eta_{2\Delta}$. However, it is not clear which microscopic mechanism is responsible for this. The authors of Ref. \cite{Fischer2023} pursued an analytical approach to solving the kinetic equations, preserving the microscopic nature of direct photon absorption. They explicitly calculate the quasiparticle-photon coupling constant \cite{Catelani2019} for $\left<n_{ph}\right>\gg1$ as the ratio between the photon absorption and the normalized power dissipation in the resonator,
\begin{equation}\label{eq:cphotqp}
	c_{phot}^{qp}=\frac{P_{abs}^{qp}}{\pi P_{int}}\frac{\sigma_N/\sigma_1}{2\hbar N_0 V} = \frac{1}{Q_i^{qp}}\frac{\sigma_N/\sigma_1}{2\hbar N_0 V}\approx\frac{\alpha_k\omega_0}{2\pi N_0 V \Delta},
\end{equation}
where the last approximation is valid for, $k_BT_{eff}\ll\Delta$, which introduces a maximum error of 7\% for our measurements. With $c_{phot}^{qp}$ known explicitly, they derive a temperature $(k_BT_*)^3/(k_B\Delta^2)$, that separates the redistribution from the quasiparticle creation regime when compared to the bath temperature. $k_BT_*$ characterizes the width of the quasiparticle distribution function and is given by, 
\begin{equation}
	\label{eq:Tstar}
	\frac{k_BT_*}{\Delta} = \left(\frac{105\pi}{64}\left(\frac{k_BT_c}{\Delta}\right)^3  \frac{\hbar\tau_0}{\Delta^2} c_{phot}^{qp} P_{int}\right)^{1/6}.
\end{equation}
Assuming $\hbar\omega_0\ll k_BT_* \lesssim \Delta$ and that the non-equilibrium phonons with energies below $2\Delta$ have little effect on the quasiparticle distribution (both valid for Al at the microwave powers considered here), they come to an explicit expression for the excess quasiparticle number at low temperatures, 
\begin{align}\label{eq:nqpsatFC}
	\begin{split}
		n_{qp}^{ex} = 0.42 &\frac{\tau_{esc}}{\tau_{pb}} N_0 \Delta \left(\frac{k_BT_*}{\Delta}\right)^{9/2}\\
		& \times \exp\left[-\sqrt{\frac{14}{5}}\left(\frac{k_BT_*}{\Delta}\right)^{-3}\right].
	\end{split}
\end{align}
This expression depends on experimental parameter $P_{int}$ as opposed \cref{eq:nqpex}, which depends on the estimated parameter $P_{abs}^{qp}$. We compare \cref{eq:nqpsatFC} to the data in \cref{fig:nqpsat}(b), without fit parameters. The variable on the x-axis combines the resonator-dependent variables in \cref{eq:Tstar,eq:cphotqp}.\\
We observe a much higher excess quasiparticle density at low microwave power densities than predicted by \cref{eq:nqpsatFC}. A similar discrepancy is observed in Ref. \cite{Fischer2023}, when comparing $Q_i(P_{int})$ to the measurements of Ref. \cite{deVisser2014b}. This suggests that the quasiparticle generation at low temperature and readout powers involves a different mechanism than direct photon absorption. From \cref{fig:nqpsat}(a) we know that such a mechanism should be readout power dependent, with an efficiency of only $4\times10^{-4}$ with respect to the absorbed power.\\
The authors of Ref. \cite{Fischer2024a} proposed pair-breaking photons generated by the signal generator as additional mechanism. In \cref{ap:100GHz}, we exclude this hypothesis by measuring the attenuation of several microwave components in the setup at frequencies above the gap frequency of the Al, which results in a total attenuation on the order of $-300~\text{dB}$. We also exclude pair-breaking photons radiated from other parts of the microwave chain. Therefore, the pair-breaking mechanism is likely to be related to the microscopics of the resonator.\\
To identify possible pair-breaking mechanisms in the resonator, we show the measured $Q_i$, normalized to quasiparticle density and resonator properties, in the inset of \cref{fig:nqpsat}(b). The solid black line is the prediction of Ref. \cite{Fischer2023} for $Q_i^{qp}$ in the quasiparticle creation regime and the dashed line is for the redistribution regime. The expressions are given in \cref{ap:Qi}. Both these predictions are higher than the measured $Q_i$, which implies that the $Q_i$ is not limited by quasiparticles (i.e. $Q_i\neq Q_i^{qp}$), but by another dissipation mechanism. This could for example be TLS loss in the inductor, which would locally heat the Al and, indirectly, generate quasiparticles. Further study on different quasiparticle generation mechanisms, both theoretical and experimental, is needed to identify such a mechanism.\\
Apart from this, the kinetic inductance nonlinearity could also play a role. Effects of the condensate are more likely to dominate at low quasiparticle densities \cite{Klapwijk2020}. Physically, the microwave drive broadens the coherence peak and creates an exponential-like tail in the density of states \cite{Semenov2016,Semenov2020a}. This causes the sharply peaked non-equilibrium distribution function to be broadened as well. A theory incorporating both the non-equilibrium density of states and redistribution effects is still lacking.

\section{Conclusion}
To conclude, we have shown that microwave readout power generates a higher excess quasiparticle density in resonators with a smaller active volume, which scales as $n_{qp}^{ex}\propto\sqrt{P_{int}/(Q_iV)}$. The generated quasiparticles limit the sensitivity of MKIDs, which can only be mitigated by simultaneously reducing $\sqrt{V\omega_0}/(Q\alpha_k)$ in the MKID design and the TLS noise. The microscopic mechanism for the quasiparticle generation is unknown, but acts local at the active resonator volume. A complete microscopic understanding needs a treatment of the non-equilibrium effects of multiple microwave power absorption channels.

\section*{Acknowledgements}
P.J.d.V. acknowledges support by the Netherlands Organisation for Scientific Research NWO (Veni Grant No. 639.041.750 and Projectruimte 680-91-127). J.J.A.B. acknowledges support by the European Research Council ERC (Consolidator Grant No. 648135 MOSAIC).\\

\section*{Author Contributions}
J.J.A.B. designed the devices, which V.M. and D.J.T. fabricated. J.B. performed the measurements and S.A.H.d.R. analyzed the data and wrote the manuscript. P.J.d.V. and J.J.A.B. supervised the entire process, including reviewing the manuscript.
\section*{Data availability}
All data and analysis scripts presented and used in this work are openly available \footnote[3]{\href{https://doi.org/10.5281/zenodo.14999891}{10.5281/zenodo.14999891}}.
\appendix
\crefalias{section}{appendix}
\section{Quasiparticle sensitivity and photon interaction}\label{ap:QPsens}
Since we probe a hybrid NbTiN-Al superconducting resonator, we are in principle sensitive to quasiparticle density changes in both the NbTiN and Al sections. In this section we will show, however, that we are only sensitive to quasiparticles in the Al section. Additionally, we will show that the photons in the resonator effectively only interact with the quasiparticle in the Al. This justifies the simplification to regard the resonator as a pure Al resonator, with a volume equal to the volume of the Al inductive strip.

\begin{figure*}
	\includegraphics[width=\textwidth]{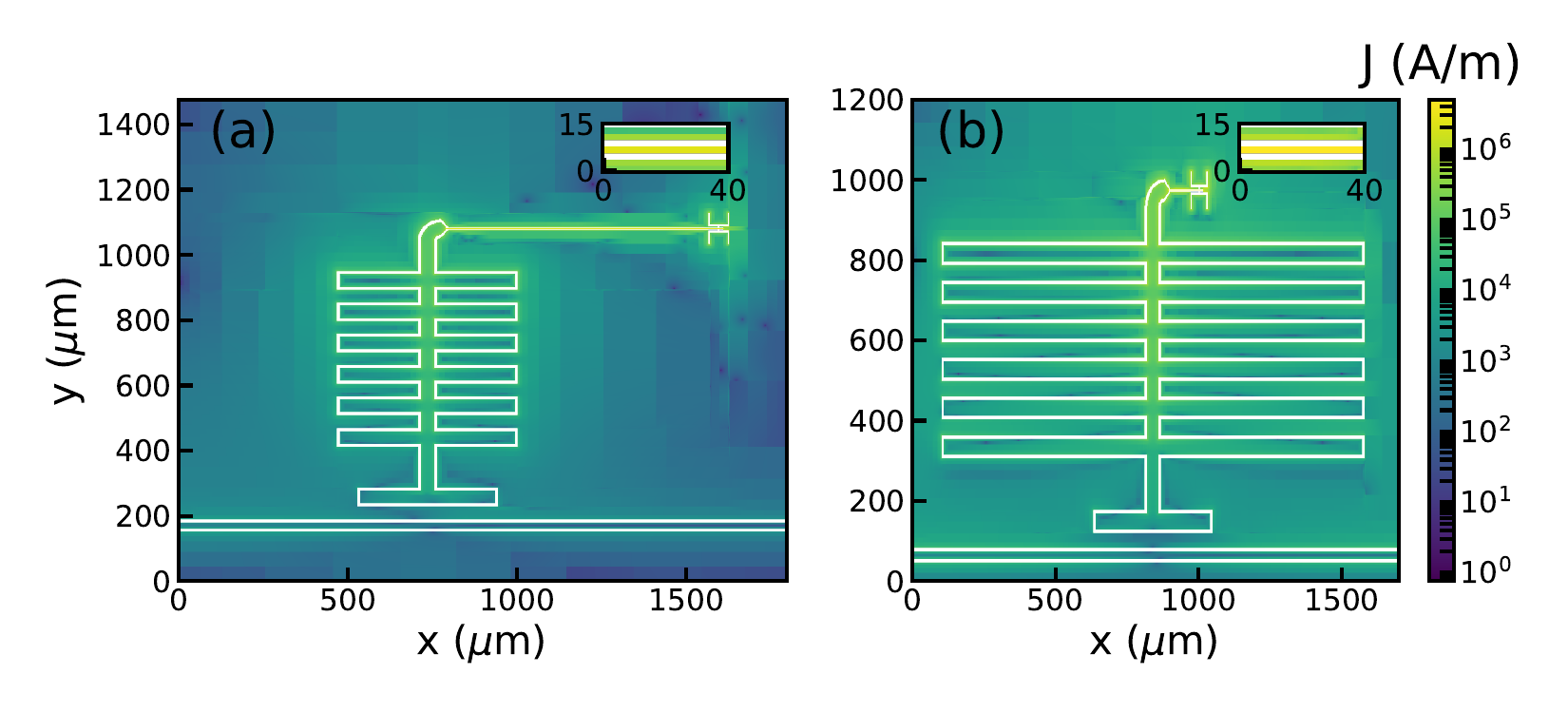}
	\caption{\label{fig:current} Current density at resonance from a simulation using SONNET \cite{Rautio2025} for the longest (a) and shortest (b) Al sections. The insets show a zoom-in of the inductive section. The colorbar relates to both panels.}
\end{figure*}

To lowest order in temperature and resonance frequency ($k_BT\ll\hbar\omega\ll\Delta$) the quasiparticle sensitivity of the amplitude, $\delta A$ and phase $\theta$ for thin films (with a thickness much smaller than the penetration depth) is proportional to \cite{deVisser2014,Gao2008} \footnote[4]{In the opposite limit, $\hbar\omega\ll k_BT \ll \Delta$, the first proportionality is slightly adjusted, but the second proportionality (keeping the material parameters only) still holds.}, 
\begin{equation} \label{eq:sensitivity}
	\frac{d(\delta A, \theta)}{dn_{qp}} \propto \frac{\alpha_k Q}{\pi N_0\Delta}\sqrt{\frac{2\Delta}{\hbar\omega}}\propto\frac{\alpha_k}{N_0\sqrt{\Delta}}.
\end{equation}
In the last expression, we only kept the material dependent parameters. The denominator, $N_0\sqrt{\Delta}$, is a factor $\sim7.0$ bigger for NbTiN, since $\Delta\approx1.76k_BT_c$ and $N_0^{NbTiN}= 3.7\times10^4 \mu\text{eV}^{-1}\mu\text{m}^{-3}$ \cite{Sidorova2021} and $N_0^{Al}= 1.7\times10^4 \mu\text{eV}^{-1}\mu\text{m}^{-3}$ \cite{Kaplan1976}. The kinetic induction fraction, $\alpha_k$, can be obtained from simulation in the following way. We first simulate the full resonator and find the resonance frequency, $f_{res}$. We set the sheet kinetic inductance to their nominal values, $L_{k,s}=\hbar R_N/(\pi\Delta)$ \cite{Mattis1958}, where $R_N$ is the normal state resistance. Then, we set the sheet kinetic inductance of the NbTiN or Al to 0 and find the resonance frequency for that case, $f_{res}^0$. We obtain the kinetic induction fraction is by $\alpha_k = 1 - (f_{res}/f_{res}^0)^2$. We simulate the resonators with longest and shortest Al sections with SONNET \cite{Rautio2025}. The resulting current density (with nominal kinetic inductance) is plotted in \cref{fig:current}. We find $\alpha^{NbTiN} = 0.12$ and $\alpha_k^{Al}=0.094 $ for the long Al resonator and $\alpha_k^{NbTiN} = 0.11 $ and $\alpha_k^{Al}=0.024$ for the short one. These Al kinetic induction fractions are consistent with the measured values from the change in resonance frequency \cite{Gao2006}, which range from 2 to 8 \% for the short and long Al section resonators, respectively. Comparing the NbTiN and Al values, the kinetic induction fraction of NbTiN is at maximum a factor $\sim 5$ higher. The total sensitivity (\cref{eq:sensitivity}) is a factor $5.5$ to $1.5$ higher for Al.\\
Furthermore, the sensitivity to the quasiparticle density is dependent on the squared current at that location $I^2(x)$ \cite{Gao2008a,Mazin2009}. To take this into account, we square the simulated current density from SONNET (\cref{fig:current}) and sum the NbTiN and Al parts individually. The squared current density is a factor 22 higher in the long Al section and a factor 3.2 higher in the short one, compared to the current density in the NbTiN. Thus, we are more sensitive to quasiparticle density changes in the Al than the NbTiN, by a factor 130 for the long Al section resonator and a factor 4.9 for the short one.

Within the Al wire, the current will be non-uniform over the wire length, since the wire has some capacitance in addition to the inductance and therefore electrical length. When we assume a sinusoidal current distribution over the wire, we can determine an effective quasiparticle volume as, 
\begin{equation}
	\begin{aligned}
		\frac{V_{eff}}{V}&=\frac{v_{ph}}{\omega_0 L}\int_{\pi/2 - \omega_0 L /v_{ph}}^{\pi/2} \sin^2(x) dx \\
		&= \frac{1}{2} + \frac{\sin(2\omega_0L/v_{ph})}{4\omega_0L/v_{ph}},
	\end{aligned}
\end{equation}
where $L$ is the inductor length, $\omega_0=2\pi f_0$ is the angular resonance frequency and $v_{ph}$ is the phase velocity of the hybrid NbTiN-Al co-planar waveguide (CPW) transmission line. The integration limits go from the phase at the NbTiN connection to the short (with a current maximum, so a phase of $\pi/2$). We obtain $v_{ph}$ from a SONNET simulation of just the inductive CPW line. For the resonator with the longest Al section, this results in a fraction $V_{eff}/V\approx 0.95$. This is consistent with the simulation shown in \cref{fig:current}(a), where the current density squared on the left side of the Al strip is $\sim95\%$ of the current on the right side. Therefore, we ignore this effect and assume that the current in the Al wire is uniform, such that we are equally sensitive to all quasiparticles in the Al.

\begin{figure*}
	\includegraphics[width=.7\textwidth]{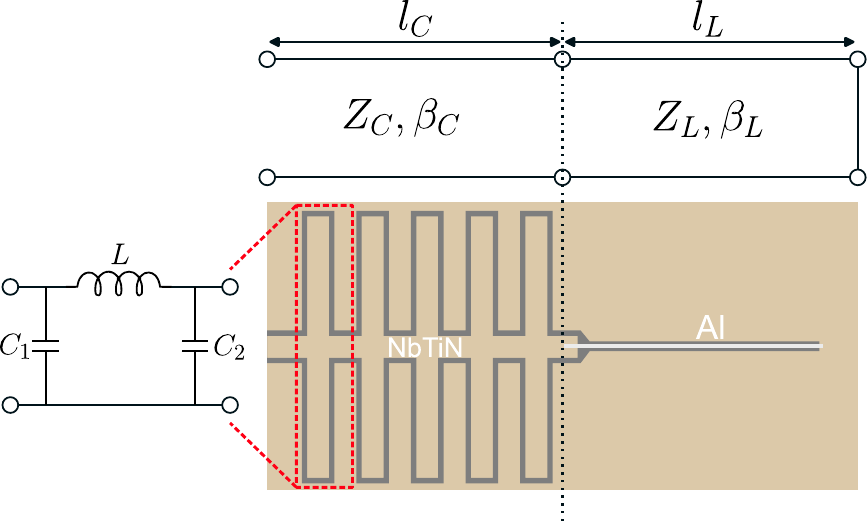}
	\caption{\label{fig:hybrid} Transmission line model for the hybrid NbTiN-Al resonator. The NbTiN capacitive and Al inductive parts are modeled by a transmission line with lengths $l_C$ and $l_L$, characteristic impedances, $Z_C$ and $Z_L$, and phase constants, $\beta_C=\omega/v_{ph}^C$ and $\beta_L=\omega/v_{ph}^L$. The inductive part is shorted to make a quarter-wave resonator. The physical structure of the transmission lines is shown below. The characteristic impedances and phase velocities are obtained from a simulation in SONNET. For the inductive Al line this results in $Z_L=83~\Omega$ and $v_{ph}^L = 1.0\times10^8~\text{m/s}$. For the NbTiN capacitive part this is done by simulating one element (highlighted in the dashed red line) and extracting the PI-model parameters as shown on the left. This way, $Z_C=\sqrt{L/(C_1+C_2)}$ and $v_{ph}=l'/\sqrt{L(C_1+C_2)}$, where $l'$ is the length of the element. This results in $Z_C=17 - 25~\Omega$ and $v_{ph}^C=2.5 - 4.0 \times10^7~\text{m/s}$, varying with capacitive finger length, see \cref{fig:current}.}
\end{figure*}

Additionally, the photons in the resonator will interact mainly with the quasiparticles in the Al. The relative photon occupation in the NbTiN capacitive part and the Al inductive part can be estimated by considering the resonator as two connected transmission lines (\cref{fig:hybrid}). We disregard the effect of the coupler, since the coupling is weak ($Q_c>30.000$). The ratio of internal power is given by $P_{int}^C/P_{int}^L = Z_L |V_C^+|^2/(Z_C |V_L^+|^2)$ \cite{Pozar2011,deVisser2014}, with $V_C^+$ and $V_L^+$ the voltage wave amplitudes in the capacitive and inductive part respectively. The voltage on the connecting point of the two transmission lines (black dashed line in \cref{fig:hybrid}) must be the same, $2 V_C^+ \cos(\beta_Cl_C) = -2j V_L^+\sin(\beta_L l_L)$. On resonance, the sum of the input impedance from the dashed line looking to the left and looking to the right, should be 0 (transverse resonance technique \cite{Pozar2011}), $j Z_C / \tan(\beta_C l_C) = j Z_L \tan(\beta_L l_L)$. Combining these two equations gives,
\begin{equation}
	\frac{P_{int}^C}{P_{int}^L} = \frac{Z_C}{Z_L}\left(\frac{\cos(\beta_L l_L)}{\sin(\beta_C l_C)}\right)^2\approx\frac{Z_C}{Z_L}.
\end{equation}
The last approximation holds if most of the electrical length is in the capacitive section. This is the case for the resonators considered here, since for the longest Al section, $(\cos(\beta_L l_L)/\sin(\beta_C l_C))^2\approx 1.2$. This results in a fraction ${P_{int}^C}/{P_{int}^L}\approx 0.24$, meaning that most of the photon occupation is in the inductive Al section.\\
Furthermore, the quasiparticle-photon interaction is much stronger in the Al section. As mentioned in the main text, the interaction strength can be estimated by, $c_{phot}^{qp}\approx \alpha_k\omega_0/(2\pi N_0 \Delta V)$ \cite{Fischer2023}. This expression has the same dependence on $\alpha_k$, $N_0$ and $\Delta$ as \cref{eq:sensitivity}, which slightly favors interaction with the Al quasiparticles. However, there is an additional factor $V$ in the expression for $c_{phot}^{qp}$. This is the volume that the quasiparticles occupy, which is much larger for NbTiN (see \cref{fig:current}). Therefore, a photon is much more likely to interact with a quasiparticle in the Al than with a quasiparticle in the NbTiN. \\
These two conclusions - the photon occupation is larger and the photon-quasiparticle interaction is stronger in the Al section compared to the NbTiN section - justify the simplification of considering the resonator as purely Al when applying the photon-quasiparticle physics described in Refs. \cite{Goldie2012,Fischer2023,Fischer2024a}.
\section{Bifurcation power estimation}\label{ap:Pbif}
When a superconducting microwave resonator is driven above a certain power, the transmission curve becomes hysteretic \cite{Swenson2013,Zmuidzinas2012}. A resonator in this state is called to be bifurcated and the internal power at which this start to happen we call the bifurcation power or $P_{int}^{bif}$.  \\
This behaviour can be modelled as a classical Duffing oscillator, with a nonlinearity parameter $a$ that describes the resonance frequency shift in number of line widths when driving at resonance. For $a>4\sqrt{3}/9=0.77$ the resonator is bifurcated \cite{Swenson2013,Zmuidzinas2012}.\\
For the kinetic inductance nonlinearity, $a$ depends on the ratio of the current density in the resonator over the critical current density in the superconductor: $j_0/j_c$ where $j_0$ is the maximum current density in one cycle. We will first find $j_0$ as a function of internal microwave power, $P_{int}$. \\
The internal power is given by \cite{Gao2008a}, 
\begin{equation}\label{Seq:Pint}
	P_{int} = \frac{\omega_0E_{res}}{4\pi m}=\frac{1}{2\pi m}\frac{Q^2}{Q_c}P_{read},
\end{equation}
with $m=1/4$ for a quarter-wave resonator (and lumped-element resonator) and $m=1/2$ for a half-wave. $\omega_0$ is the angular resonance frequency, $Q=(1/Q_c + 1/Q_i)^{-1}$ is the loaded quality factor with $Q_c$ the coupling quality factor and $Q_i$ the internal quality factor. $E_{res}$ is the energy stored in the resonator, which can be expressed as, 
\begin{equation}\label{Seq:Eres}
	E_{res} = \left<n_{ph}\right>\hbar\omega_0 = \frac{1}{2}L I_0^2,
\end{equation}
with $\left<n_{ph}\right>$ the average number of photons in the resonator, $\hbar$ the reduced Planck constant, $L$ the total inductance and $I_0$ the maximum current in the inductor over one cycle. In this equation, we assumed the current to be constant over the inductor wire and that all inductive energy in the resonator can be described by a single lumped element inductor with inductance $L$ (see previous section). If we would consider a completely distributed resonator, i.e. a transmission line of finite length, there would be an additional factor $1/2$ in \cref{Seq:Eres}. For now, we consider a quarter-wave resonator and set $m=1/4$.\\
Combining \cref{Seq:Pint,Seq:Eres}, we see,
\begin{equation}
	j_0 = \sqrt{\frac{2\pi P_{int}}{L\omega_0 (dw)^2}},
\end{equation}
where $j_0=I_0/(dw)$ is the current density, $d$ is the wire thickness and $w$ is the wire width. The total inductance can be calculated via,
\begin{equation}
	L = \frac{L_k}{\alpha_k}=\frac{L_{k,s}}{\alpha_k}\frac{l}{w},
\end{equation}
with $l$ the wire length, $\alpha_k$ the kinetic inductance fraction, $L_k$ the total kinetic inductance and $L_{k,s}$ the sheet kinetic inductance. In the thin film, local limit we have $L_{k,s} = 1/(d\sigma_2\omega_0)$ \cite{Gao2008a} and at low temperatures, $k_BT\ll\Delta$, the imaginary part of the complex conductivity equals, $\sigma_2 = \pi\Delta/(\hbar\omega_0\rho_N)$ \cite{Mattis1958}, with $\rho_N$ the normal state resistivity. Putting this together, we find, 
\begin{equation}\label{Seq:j0}
	j_0 = \sqrt{\frac{4\pi^2}{\rho_N}\frac{\Delta}{\hbar\omega_0}\frac{\alpha_k P_{int}}{V}}.
\end{equation}

The critical current density is given by \cite{Anthore2003,Clem2012} \footnote[2]{This equation is derived with via the Usadel equations, which holds for $T\ll T_c$ as opposed to the derivations using the Ginzburg-Landau equations, $j_c=1.54 j_*$ \cite{Clem2012}}, 
\begin{equation}\label{Seq:jc}
	j_c = 0.59 \sqrt{\frac{\pi\Delta^3N_0}{\hbar\rho_N}}=0.59 j_*,
\end{equation}
where $j_*$ is defined as the square-root term by convention. Because we assume that the current density is uniform over the wire, this critical current is given by the depairing current. If the wire would not be straight, but would have constrictions, sharp turns, or would be wider than the penetration depth, the critical current would be reduced \cite{Clem2011,Clem2012}.\\
To first order, the kinetic inductance changes with (uniform) current as, $L_k(j) = L_k(0) (1 + 0.069(j/j_c)^2)$ \cite{Semenov2016, Semenov2020} under AC field, which coincides with the DC case when we take $j_{rms} = j_0/\sqrt{2} \rightarrow j_{DC}$ \cite{Anthore2003, Semenov2020}. This results in a fractional frequency shift compared to the zero current case \footnote[1]{Ref. \cite{Semenov2016} omitted a factor $\alpha_k$ as pointed out by Ref. \cite{Fischer2024a}. However, an additional factor 0.16 is omitted in the derivation of $\delta\omega/\omega_0$ in Ref. \cite{Semenov2016}. Since the resonator studied there have a $\alpha_k$ close to 0.16, these errors approximately cancel. \cref{Seq:dww0} is consistent with Ref. \cite{Semenov2016}, when these two errors are taken into account and considering that $P_0$ in Ref. \cite{Semenov2016} can also be expressed as $P_0=2N_0 V \Delta^2 \omega_0/\alpha_k$ for a distributed, half-wave resonator.},
\begin{equation}
	\label{Seq:dww0}
	\begin{aligned}
		\left.\frac{\delta\omega}{\omega_0}\right|_{L_k}&=-\frac{\alpha_k}{2}\frac{\delta L_k}{L_k(0)} \\
		&= - 0.035 \alpha_k \left(\frac{j_0}{j_c}\right)^2 \\
		&= - 1.2 \frac{\alpha_k^2 P_{int}}{N_0 V\Delta^2\omega_0}.
	\end{aligned}
\end{equation}

\begin{figure*}
	\includegraphics[width=.7\textwidth]{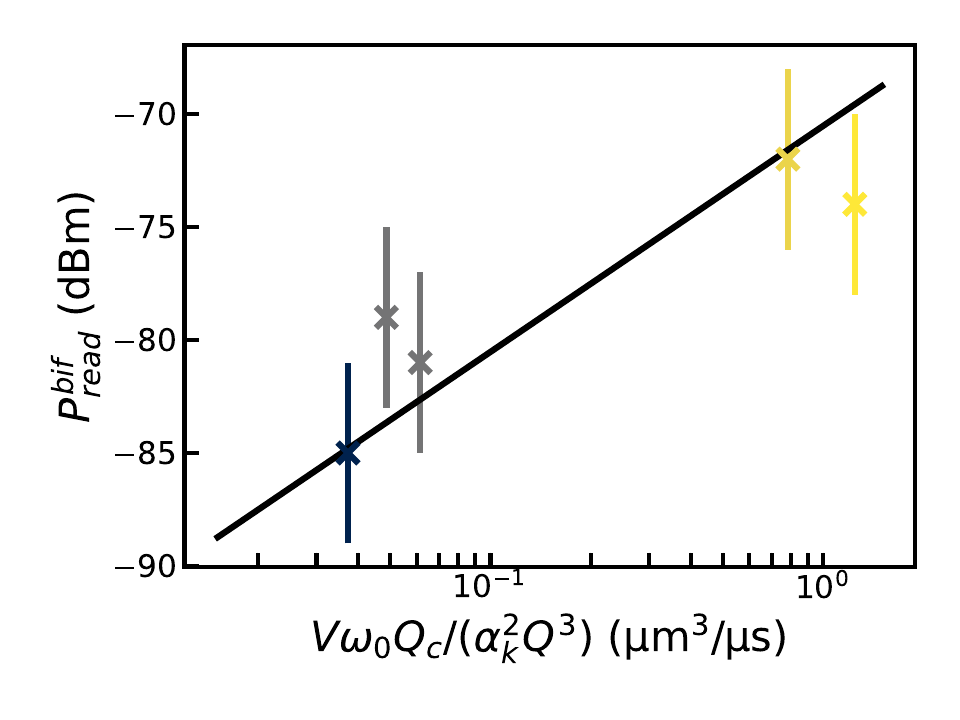}
	\caption{\label{Sfig:Pbif} Verification of \cref{Seq:Preadbif} with the resonators of the main text. $P_{read}^{bif}$ is measured by visually checking the transmission curves at $T=100~\text{mK}$ with 4 dB steps in read power, which is indicated by the error bars. The black line is \cref{Seq:Preadbif} without fit parameters, with $Q_c$ and $Q$ from a Lorentzian fit to the same transmission curves. $\alpha_k$ is obtained from transmission curves at higher bath temperatures \cite{Gao2006}.}
\end{figure*}

Finally, bifurcation occurs when $\delta\omega/\omega_0 = - 0.77 / Q$ \cite{Zmuidzinas2012,Swenson2013}, which results in 
\begin{equation}
	\label{Seq:Pbif}
	P_{int}^{bif} = 0.64 N_0\Delta^2 \frac{V \omega_0}{\alpha_k^2 Q},
\end{equation}
and for the microwave power on the read out line (see \cref{Seq:Pint}),
\begin{equation}
	\label{Seq:Preadbif}
	P_{read}^{bif} = N_0\Delta^2 \frac{V \omega_0 Q_c}{\alpha_k^2 Q^3}.
\end{equation}
This equation agrees closely with the resonator data from the main text, see \cref{Sfig:Pbif}.\\

\subsection{Quasiparticle non-linearity}
Apart from the kinetic inductance non-linearity discussed above, the quasiparticle redistribution effects of read power also introduce a non-linearity in resonance frequency \cite{Goldie2012,Fischer2023,Fischer2024a}. We can estimate this via the excess quasiparticle density, as $\delta\omega/\omega_0=-\alpha_k n_{qp}^{ex}/(4N_0\Delta)$. The corrections for the reduced gap energy and quasiparticle distribution shape for this expression are of order unity \cite{Fischer2024a}; approximately a factor 1.7 for the powers close to bifurcation in the resonators considered here.\\
We can take $n_{qp}^{ex}$ from \cref{eq:nqpex} from the main text, if we know $Q_i$ and $\Sigma_s$ in terms of superconductor and resonator properties. We can approximate,
\begin{equation}\label{Seq:Qiqp}
	Q_i^{qp} = \frac{\sigma_2}{\alpha_k \sigma_1} \approx \frac{2\pi N_0\Delta }{\alpha_k n_{qp}}\sqrt{\frac{\hbar\omega_0}{2\Delta}},
\end{equation}
which holds for $k_BT\ll(\Delta,\hbar\omega_0)$ \cite{Mattis1958}.\\
If we assume that for $\Sigma_s$ the power flow from quasiparticle to phonon system is via recombination only, we can equate,
\begin{equation}
	\frac{P_{abs}}{V} = \frac{\Delta n_{qp}}{\eta_{2\Delta}\tau_{qp}^*} = \frac{\Sigma_s^{rec}\tau_0(k_BT_c)^3}{16\pi\eta_{2\Delta}N_0\Delta^3k_B}\frac{n_{qp}}{\tau_{qp}^*},
\end{equation}
where the last expression is from Ref. \cite{Goldie2012}. $\eta_{2\Delta}$ is the faction of pair-breaking phonons in the non-equilibrium phonon distribution function \cite{Goldie2012}. Solving for $\Sigma_s^{rec}$ gives,
\begin{equation}
	\Sigma_s^{rec} = 2\pi k_B (2N_0\Delta)^2R,
\end{equation}
where $R=2\Delta^2/((k_BT_c)^3N_0\tau_0)$. For Al, this results in $2.7\times10^{10}~\text{W/m}^3/\text{K}$, which is close to the value of $3.4\times10^{10}~\text{W/m}^3/\text{K}$ from numerical calculations \cite{Goldie2012}. Combining the analytical expression for $\Sigma_s^{rec}$, $Q_i^{qp}$ and \cref{eq:nqpex} results in,
\begin{equation}
	\label{Seq:nqpex}
	n_{qp}^{ex} = \frac{\alpha_k P_{int}}{2N_0\Delta V}\sqrt{\frac{2\Delta}{\hbar\omega_0}} \frac{\eta_{2\Delta}}{\pi\Delta\bar{R}}.
\end{equation}
Here, $\bar{R}=R/(1 + \tau_{esc}/\tau_{pb})$, with $\tau_{esc}$ the phonon escape time, $\tau_{pb}$ the phonon pair-breaking time and $\tau_0$ the electron-phonon characteristic interaction time.\\
Combining these expressions results in a resonance frequency shift due to the quasiparticle nonlinearity, 
\begin{equation}
	\label{Seq:dww0_qp}
	\left.\frac{\delta\omega}{\omega_0}\right|_{qp} = - \frac{\alpha_k^2 P_{int}}{N_0 V \Delta^2} \frac{\eta_{2\Delta}}{8\pi\bar{R}N_0\Delta}\sqrt{\frac{2\Delta}{\hbar\omega_0}}.
\end{equation}

From \cref{Seq:dww0,Seq:dww0_qp} we thus find the ratio of the frequency shift due to kinetic inductance effects and quasiparticle redistribution effects to be,
\begin{equation}\label{Seq:dwLkdwqp}
	\frac{\left.\delta\omega\right|_{L_k}}{\left.\delta\omega\right|_{qp}} = 21 \frac{\bar{R}N_0}{\eta_{2\Delta}}\sqrt{\frac{\hbar\Delta}{\omega_0}}.
\end{equation}
When filling in $\eta_{2\Delta}=4\times10^{-4}$, the measured Al properties, $\tau_{esc}=0.35~\text{ns}$ and the resonance frequencies from the main text, we obtain $5.1-6.1$ for this ratio. This implies the two nonlinear effects are of the same order of magnitude and the measured bifurcation is caused by the kinetic inductance nonlinearity that is described by \cref{Seq:Pbif,Seq:Preadbif}.\\
In \cref{Seq:Qiqp}, we also assumed a thermal quasiparticle distribution, which seems to be the case in our measurement, as we verified in the main text. However, the quasiparticle distribution is generally not thermal when microwave photon absorption is considered \cite{deVisser2014a,Goldie2012,Fischer2023}. The non-thermal distribution increases $Q_i$, which favours the kinetic inductance nonlinearity. See Section 3 for a quantitative description of this effect.

\section{NEP limited by microwave induced excess quasiparticles}\label{ap:NEP}
We can estimate the NEP from quasiparticle fluctuations as \cite{deVisser2012}
\begin{equation}
	\text{NEP}_{\text{GR}} = \frac{2\Delta}{\eta_{pb}} \sqrt{\frac{n_{qp} V}{\tau_{qp}^*}}=\frac{2\Delta}{\eta_{pb}} n_{qp} \sqrt{\bar{R}V},
\end{equation}
where we used that, in the quasiparticle creation regime, the effective quasiparticle lifetime is given by, $\tau_{qp}^* = 1/(\bar{R}n_{qp})$ \cite{Fischer2024a}. If we use \cref{Seq:nqpex}, i.e. assume $k_BT \ll (\Delta, \hbar\omega_0)$, and assume $P_{int}^{bif}$ from \cref{Seq:Pbif} for $P_{int}$ , we arrive at
\begin{equation}
	\text{NEP}_{\text{GR}}^{bif} = 0.29 \frac{\eta_{2\Delta}}{\eta_{pb}} \sqrt{\frac{\Delta^3}{\hbar\bar{R}}} \frac{\sqrt{V\omega_0}}{\alpha_k Q}.
\end{equation}

\section{$Q_i$ when quasiparticles are redistributed}\label{ap:Qi}
Eq. (62) from \cite{Fischer2023} gives for the redistributed quasiparticle $Q_i$ in the quasiparticle creation regime, 
\begin{equation}
	\begin{aligned}
		\left.Q_i^{qp,red.}\right|_{crea} &= \frac{19.3}{\alpha_k}\frac{\Delta}{\hbar\omega_0}\frac{\tau_{pb}}{\tau_{esc}}\left(\frac{\Delta}{k_BT_*}\right)^3 e^{\sqrt{14/5}(\Delta/k_BT_*)^3} \\
		&= \frac{91.9}{\alpha_k}\frac{\Delta}{\hbar\omega_0}\frac{2 N_0 \Delta}{n_{qp}^{ex}}\left(\frac{k_BT_*}{\Delta}\right)^{3/2},
	\end{aligned}
\end{equation}
where we used \cref{eq:nqpsatFC} from the main text in the last equality. \\
In the redistribution regime, $Q_i$ is predicted to follow Eq. (63) of \cite{Fischer2023}, 
\begin{equation}
	\left.Q_i^{qp,red.}\right|_{red.} = \frac{4.1}{\alpha_k}\frac{\Delta}{\hbar\omega_0}\frac{2 N_0 \Delta}{n_{qp}}\left(\frac{k_BT_*}{\Delta}\right)^{3/2}.
\end{equation}
This equation should also hold when $n_{qp}$ is non-thermal due to another pair-breaking process, when setting $n_{qp}\rightarrow n_{qp}^{ex}$ \cite{Fischer2024a}.

\section{Pair breaking photon attenuation}\label{ap:100GHz}
\begin{figure*}
	\includegraphics[width=\textwidth]{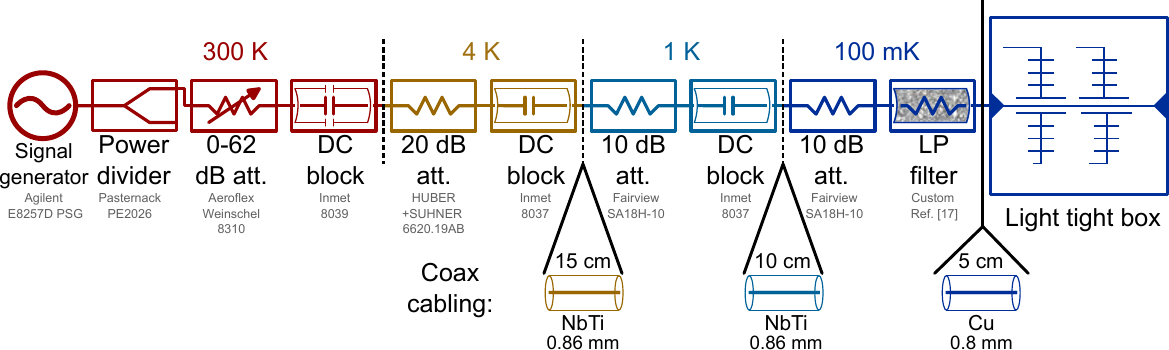}
	\caption{\label{Sfig:MWdiagram} Diagram showing the measured components in the microwave chain from signal generator to chip.}
\end{figure*}

\begin{figure*}
	\includegraphics[width=\textwidth]{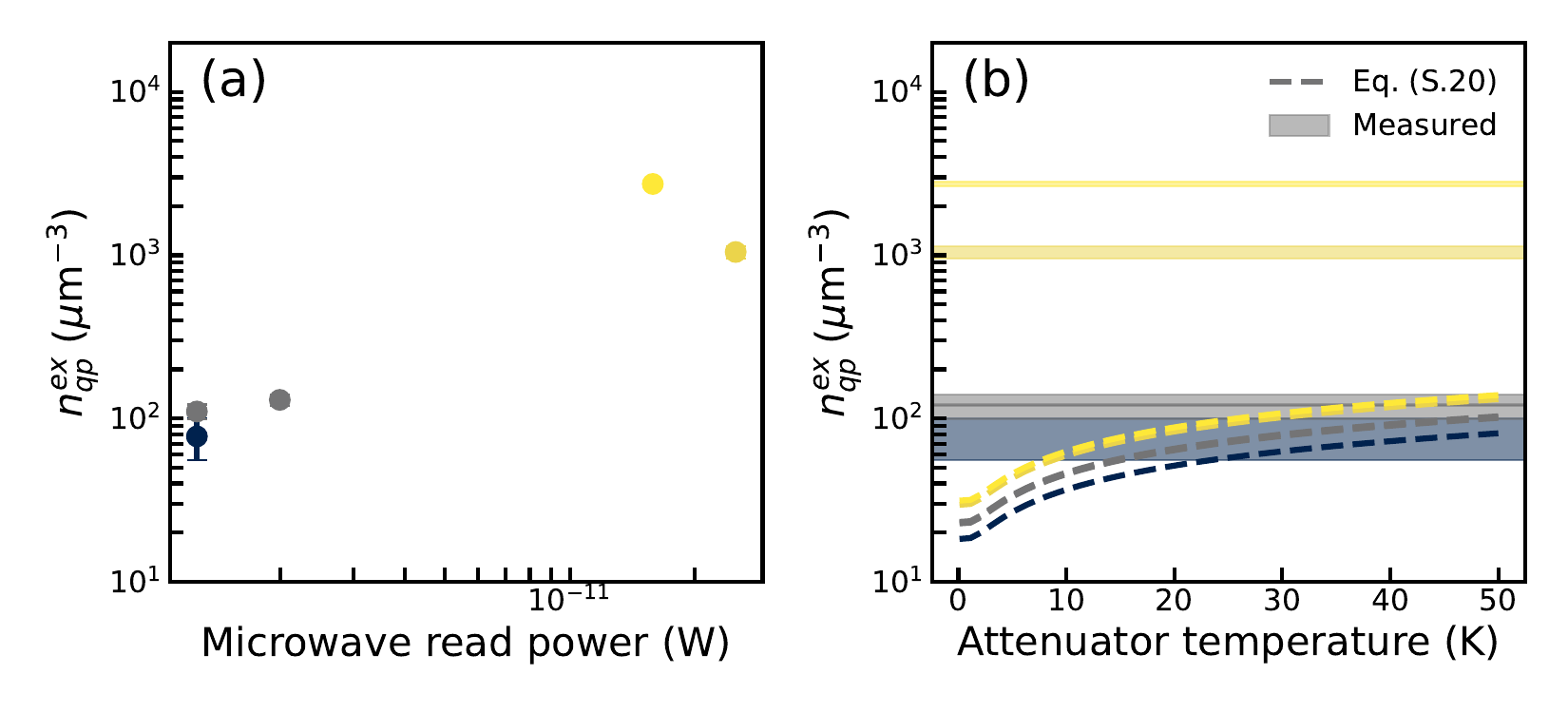}
	\caption{\label{Sfig:MWheating} Estimation for excess quasiparticle density due to read out power heating of the 10 dB attenuator at $100~\text{mK}$. (a): The measured excess quasiparticle density from \cref{fig:nqpsat}, with the used on-chip read power on the x-axis. This shows that the quasiparticle density does increase with increasing read power. (b): Estimation of the excess quasiparticle density from \cref{Seq:nqpex_phot} \cite{Fischer2024a}, when assuming a heated temperature for the 10 dB attenuator given by the x-axis. Different colours indicate the different resonators, like in \cref{Sfig:Pbif,fig:results}. The coloured areas are the measured values from (a). Since these are much larger than the predicted values (dashed lines), we conclude that this effect is too small to explain the measured excess quasiparticle densities.}
\end{figure*}

From \cref{fig:nqpsat}(b), we concluded that the measured excess quasiparticle density cannot be explained by direct microwave photon absorption by quasiparticles only. This is a similar conclusion as the authors of Ref. \cite{Fischer2023} when comparing their explicit expression for $Q_i(P_{int})$ with the data from Ref. \cite{deVisser2014b}. There must be an additional quasiparticle generation effect. \\
Since we observe that the excess quasiparticle density increases with increasing read power (\cref{Sfig:MWheating}(a)), as is observed in Ref. \cite{deVisser2014b}, this additional generation mechanism must increase with increasing read power. In Ref. \cite{Fischer2024a}, the authors propose a \textit{pair-breaking} photon occupation number that is read power dependent as this additional generation mechanism. They show that they can fit the data of Ref. \cite{deVisser2014b} when they assume a pair-breaking photon occupation in the resonator, that is a factor $10^{-9}$ of the microwave photon occupation, i.e. -90 dBc.\\
The origin of these pair-breaking photons is not clear. Two possible sources are higher harmonics of the signal generator, also mentioned in \cite{Fischer2024a}, and heating of the 10 dB attenuator on the 100 mK stage (see \cref{Sfig:MWdiagram}). We show in this section that these hypotheses cannot explain the measured quasiparticle densities.\\

\begin{figure*}
	\includegraphics[width=\textwidth]{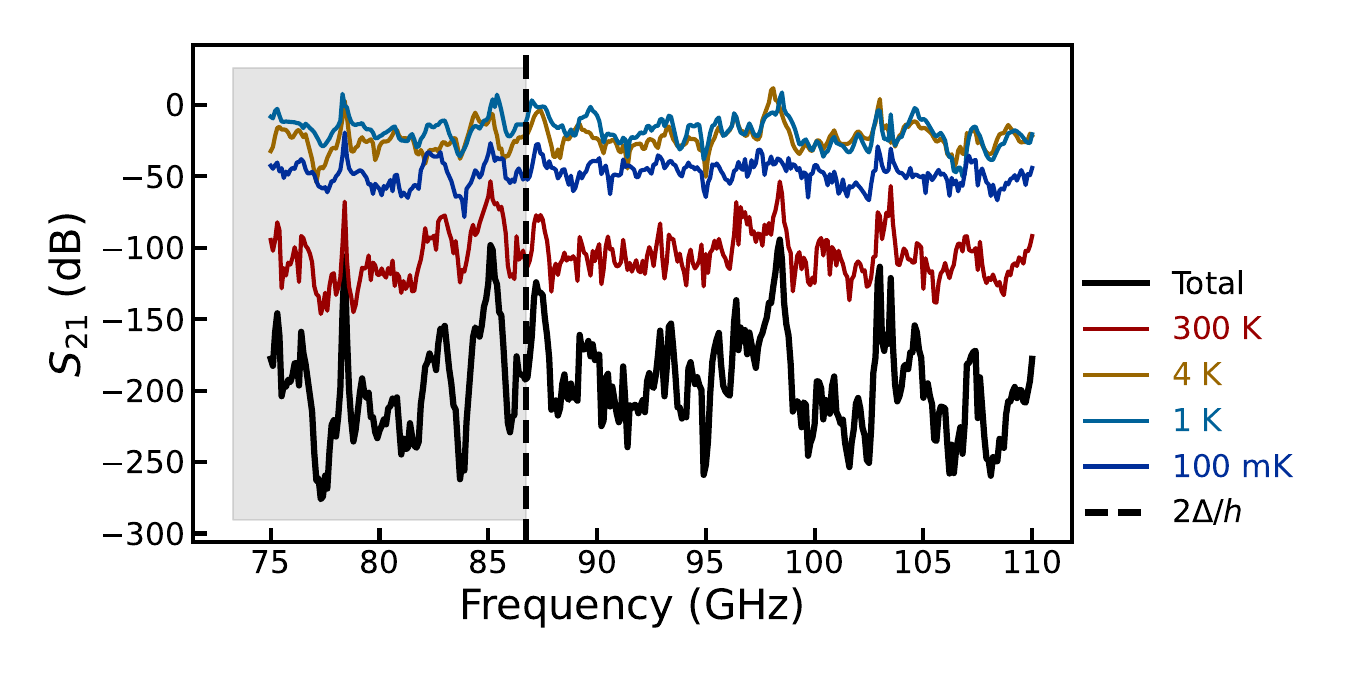}
	\caption{\label{Sfig:attmeas} Measured forward transmission of the total combined microwave chain (black) and the components of the various temperature stages, indicated by the colours. The gap frequency of the Al is indicted by the black dashed line.}
\end{figure*}

To estimate the effect of high harmonics generated by the signal generator, we measured the attenuation of pair-breaking radiation by several components in our setup, see \cref{Sfig:MWdiagram}. We used a Keysight-Agilent N5242A microwave network analyser and two Anritsu 3740A-EW modules for frequency up-conversion to the 75-110 GHz band, with adapters to go from WR-10 to SMA. We calibrated the setup with a THRU calibration, including adapters, and measured each component in \cref{Sfig:MWdiagram} individually. The combined results of these measurements are presented in \cref{Sfig:attmeas}. The main contributions to the attenuation are the variable attenuator at 300 K and the custom made low-pass powder filter \cite{Baselmans2012a} at 100 mK. We measured an attenuation of -70 dB at most frequencies for these components, which is the noise floor of the setup. The actual attenuation is therefore likely to be higher.\\
On the other hand, the measured shown in \cref{Sfig:attmeas} are performed at room temperature, while the components in the setup are cooled to the temperature specified in \cref{Sfig:MWdiagram}. We therefore disregard the losses in the superconducting NbTi wires. For the copper wire, we divide the attenuation (in dB) by the residual resistance ratio (RRR), which we assume to be 40. By doing this, we disregard losses in the dielectric of the coax cables, which might be significant at these frequencies. These corrections therefore lead to a conservative estimate of the attenuation in the wires. The attenuation of the low-pass filter is induced primarily by eddy currents in the metal powder dielectric, which is a bronze-Stycast mixture \cite{Baselmans2012a}. This attenuation has only a weak temperature dependence \cite{Milliken2007}. To account for this, we assume a RRR of bronze of 3 and divide the measured attenuation by the square root of that, as the skin depth in the metal grains scales with the square root of the resistance. These corrections are included in \cref{Sfig:attmeas}.\\
The signal generator is specified to generate less than -55 dBc in higher harmonics \cite{zotero-4699}. The average measured attenuation in the setup is -117 dBc $\pm$ 10 dB, see \cref{Sfig:attmeas}. The pair-breaking photon occupation at the chip is thus at maximum -162 dBc. This is much lower than the required -90 dBc \cite{Fischer2024a} to explain the data from Ref. \cite{deVisser2014b}.\\
For the measurement presented in the main text, we estimate the excess quasiparticle density generated by this pair-breaking photon occupation  via \cite{Fischer2024a},
\begin{equation}\label{Seq:nqpex_phot}
	n_{qp}^{ex}  = 2N_0 \Delta \sqrt{\frac{\pi}{4}\left<n_{pb}\right>c_{phot,pb}^{qp}\tau_0\frac{\xi}{\Delta}\left(\frac{k_BT_c}{\Delta}\right)^3}.
\end{equation} 
$c_{phot,pb}^{qp}=\alpha_k\hbar\omega_{pb}^2/(2\pi N_0 \Delta^2 V)$ \cite{Fischer2024a} is the quasiparticle-photon coupling constant for pair-breaking photons with energy $\hbar\omega_{pb}$. $\left<n_{pb}\right>$ is the pair-breaking photon occupation and $\xi=2\Delta - \hbar\omega_{pb}$.\\
If we take for $\left<n_{pb}\right>=\chi P_{read} / (\hbar \omega_{pb}^2)$, with $\chi=10^{-16.2}$, or $-162~\text{dB}$, the on-chip ratio of readout power and pair-breaking photons and $\hbar\omega_{pb}=2.8\Delta$ \cite{Fischer2024a}, we come to $n_{qp}^{ex}\approx 10^{-3}~\mu\text{m}^{-3}$. This is 5 orders of magnitude lower than measured at these read powers, see \cref{Sfig:MWheating}(a). We therefore exclude the signal generator as pair-breaking photon source.\\

Another possible source pair-breaking photons is radiation coming from the 10 dB attenuator at the 100 mK stage, see \cref{Sfig:MWdiagram}. We consider this the only relevant component for heating effects, since the read power must significantly heat the component to obtain the read power dependence of $n_{qp}^{ex}$ (\cref{Sfig:MWheating}(a)).\\
We estimate the on-chip pair-breaking power coming from the 10 dB attenuator as 1D black-body radiation (i.e. Johnson-Nyquist noise \cite{Nyquist1928}) that is attenuated by the low-pass powder filter,
\begin{equation}
	P_{chip} = \int_{2\Delta/h}^\infty \left[\frac{h f}{e^{hf/k_BT_{att.}} - 1}+\frac{hf}{2}\right]e^{-f/f_c} df,
\end{equation}
where $h=2\pi \hbar$ is the Planck constant, $T_{att.}$ is the temperature of the 10 dB attenuator and $f_c=10~\text{GHz}$ is the cut-off frequency of the filter \cite{Baselmans2012a}. This filter shape has been confirmed up to 10 GHz. At 80 GHz this would be a attenuation of -45 dB, where we measured it to be at least -70 dB. Thus, the filter attenuates more at higher frequencies than we assume here.\\
We calculate the average photon energy as, 
\begin{equation}
	\left<E_{ph}\right> = \left.P_{chip}\middle/\int_{2\Delta/h}^\infty \left[\frac{1}{e^{hf/k_BT} - 1}+\frac{1}{2}\right]e^{-f/f_c} df\right.,
\end{equation}
which results in $\xi=0.22$. Using $\left<n_{pb}\right>=hP_{chip}/\left<E_{ph}\right>^2$ in \cref{Seq:nqpex_phot}, we obtain the dashed lines in \cref{Sfig:MWheating}(b). The different lines correspond to the different resonators. We deem it extremely unlikely that the attenuator at 100 mK is heated to 50 K or more, since it has a stainless steel body that is connect directly to the cryostat 100 mK stage. Moreover, we here did not consider the coupling of pair-breaking photon from transmission line to the Al section of the resonator. That would give an additional attenuation of -32 dB, as estimated from a simulation of the resonator in SONNET. When taking this coupling into account, the dashed lines in \cref{Sfig:MWheating}(b) shift more than an order of magnitude downwards. We therefore conclude that also this mechanism is not the cause of the measured excess quasiparticle density.
\clearpage

\end{document}